\documentclass[10pt,a4paper,twoside,onecolumn]{article}
\usepackage[T1]{fontenc}
\usepackage[english]{babel}
\usepackage[dvips]{graphicx}
\usepackage{fancyhdr}
\usepackage{supertabular}
\usepackage{multirow}

\def\titlefont{\fontsize{12}{13}\bfseries\boldmath\selectfont\centering{}}
\def\authorfont{\footnotesize}

\let\affiliationfont\rhfont

\def\title#1{
    \vspace*{-14pt}
    \vskip 79pt
        {\centering{\titlefont #1\par}}%
    \vskip 1em
}

\def\author#1{\par
    {\centering{\authorfont#1}\par\vspace*{0.05in}}
}

\def\address#1{\par
    {\centering{\affiliationfont#1\par}}\par\vspace*{11pt}
}

\def\keywords#1{\par
    \vspace*{8pt}
    {\authorfont{\leftskip18pt\rightskip\leftskip
    \noindent{\it Keywords}\/:\ #1\par}}\vskip-12pt}
\begin{document}

\title{Exclusive electroproduction of the $\rho^+$ meson on the proton \\ @ CLAS}
\author{A. Fradi}

\address{Univ Paris-Sud, Institut de Physique Nucl\'eaire d'Orsay\\
91405 Orsay,  France\\
$^*$E-mail: fradi@ipno.in2p3.fr\\}

\begin{abstract}
We present preliminary results of the exclusive electroproduction of  $\rho^+$ on the proton at CLAS.
We discuss the interpretation of the cross sections in terms of
$t$-channel Reggeon exchanges and in terms of Generalized Parton
Distributions (GPDs) formalism. 
\end{abstract}


\section{Introduction}\label{aba:sec1}

The exclusive electroproduction of mesons on the nucleon is an important tool to better understand the
nucleon structure and, more generally, the transition between the low energy hadronic and high energy partonic
domains of the Quantum Chromodynamics (QCD) theory.
 The CEBAF (``Continous Electron Beam Accelerator Facility'') of the Jefferson Laboratory (JLab) at Newport News (USA) with
 its high intensity electron beam  and its large acceptance spectrometer CLAS \cite{clas} (CEBAF Large Acceptance Spectrometer) 
 offers a great opportunity to study the exclusive electroproduction of mesons,
 in particular those that have multi-particle decays.
 
 This proceeding presents some first results for the exclusive electroproduction of the vector mesons $\rho^+$ on the
 proton  at CLAS. These results come from  the e1-dvcs experiment: the data were collected between March and May
 2005 with a beam energy of 5.75 GeV and an integrated  luminosity of $\sim 40$ fb$^{-1}$. It is the first ever measurement of 
 cross section for the $\rho^+$ electroproduction channel. The presented results are still preliminary and more details can be found in ref. \cite{mathese}.

 \section{Data analysis}

\begin{figure}[!h]
\begin{center}
\includegraphics[width=14cm]{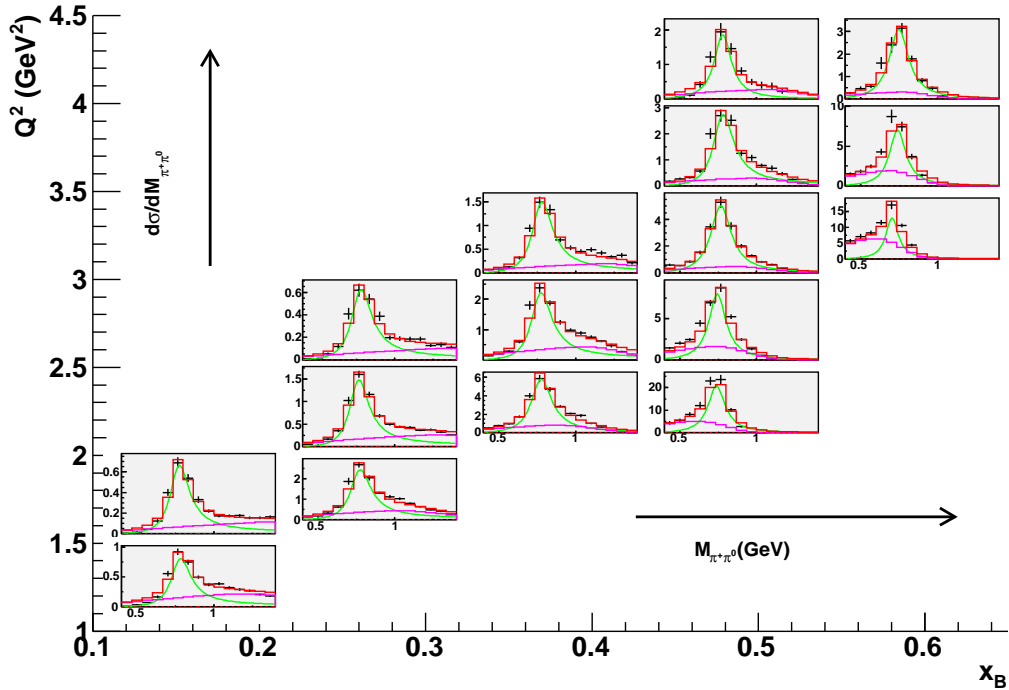}
\caption{$M_{\pi^+ \pi^0}$ acceptance-corrected distributions, showing fits for the background
subtraction. In black: experimental data; in green: Breit-Wigner of $\rho^+$; in purple: $M_{\pi^+ \pi^0}$ 
projection of the the non-resonant continuum $\gamma^* p \to n \pi^+ \pi^0$ reaction; in red: total fit result.
 In this proceeding, units are arbitrary for the vertical scale.} \label{fitQ2xB}
\end{center}
\end{figure}

The final state that is analyzed is $e p \to e n \pi^+ \pi^0$. The final state particles $e$, $\pi^+$ and $\pi^0$ were detected in
 CLAS and the exclusive process was identified by selecting 
a neutron  with the missing mass technique. Figure \ref{fitQ2xB} shows the
acceptance-corrected invariant mass $M_{\pi^+ \pi^0}$  distributions for each
  $(Q^2,x_B)$ bin. One sees clearly the $\rho^{+}$ peaks around 775 MeV. They  are, however,
  sitting on top of a non-negligeable non-resonant, two-pion background.  We have used two contributions to fit these spectra: a skewed Breit-Wigner distribution to describe
  the resonant structure of the $\rho^+$ and a parametrization
of the non-resonant two-pion continuum  determined from simulations (mostly $e p \to e n \pi^+ \pi^0$ phase space). 
In order to extract the total cross section of the reaction  $\gamma^{*}  p \to  n  \rho^{+}$, we have calculated for each 
    $(Q^2,x_B)$ bin the area of the Breit-Wigner (green curve in Fig. \ref{fitQ2xB}), the distributions being already normalized 
    and corrected for the acceptance. A separation of the longitudinal cross sections from  the transverse ones was performed  by analysing the pion 
decay angles of the $\rho^+$ and relying on the SCHC ($s$-channel helicity conservation), which was checked experimentally.
    
\section{Interpretation}\label{theory}

\begin{figure}
 \begin{center}
\includegraphics[height=3cm]{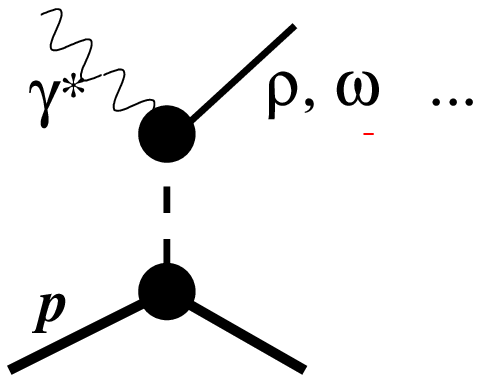}\hfill
\includegraphics[height=3cm]{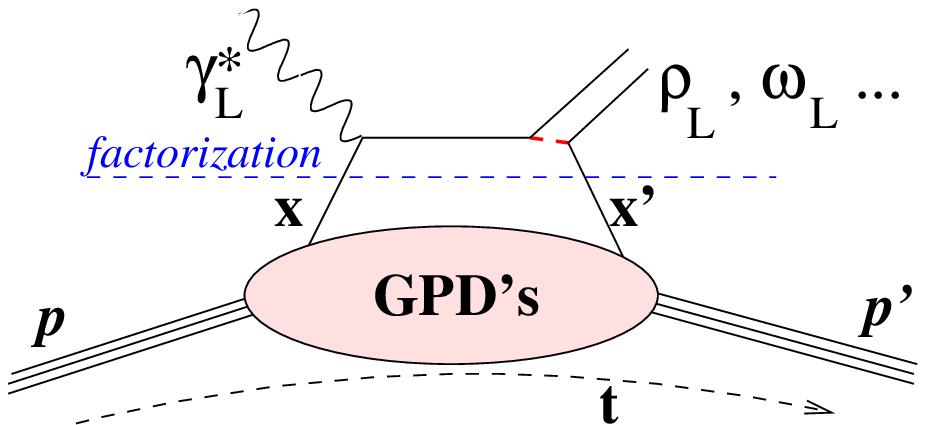}
\caption{Shematic representations of the Reggeon $t$-channel exchange (left) and of the handbag diagram (right) for exclusive vector
  meson electroproduction.}\label{reggeGPDs}
\end{center}
\end{figure}

 We will compare in the following the extracted cross sections of the exclusive
 electroproduction of the $\rho^+$ with two theoretical approaches: on the one hand, the hadronic approach based on Regge
 theory and meson trajectory exchanges in the $t$-channel and on the other hand, the partonic approach based on the handbag diagram and GPDs. 
 These two approaches are illustrated in Fig. \ref{reggeGPDs}.
This comparison will be useful in order to  better understand the
 domain of validity of these two approaches and constrain their inputs.
 
   \subsection{The Regge ``hadronic'' approach}

The Regge approach is most appropriate at low $Q^2$ above the resonance region and at forward angles (where the cross 
section is largest). It consists of exchanges of meson ``trajectories'' in the $t$-channel.
In the following, we will use the ``JML'' acronym to refer to the specific model developed by J.-M. Laget \cite{JML}.

For the $\rho^+$ channel, the exchange trajectories are those of the  $\pi^+$ and of the $\rho^+$.
 The free parameters in this model are the coupling constants at the hadronic vertices
(most of them being well constrained) and the mass scales of the electromagnetic form factors.

\begin{figure}[!h]
\begin{center}
\includegraphics[width=14cm]{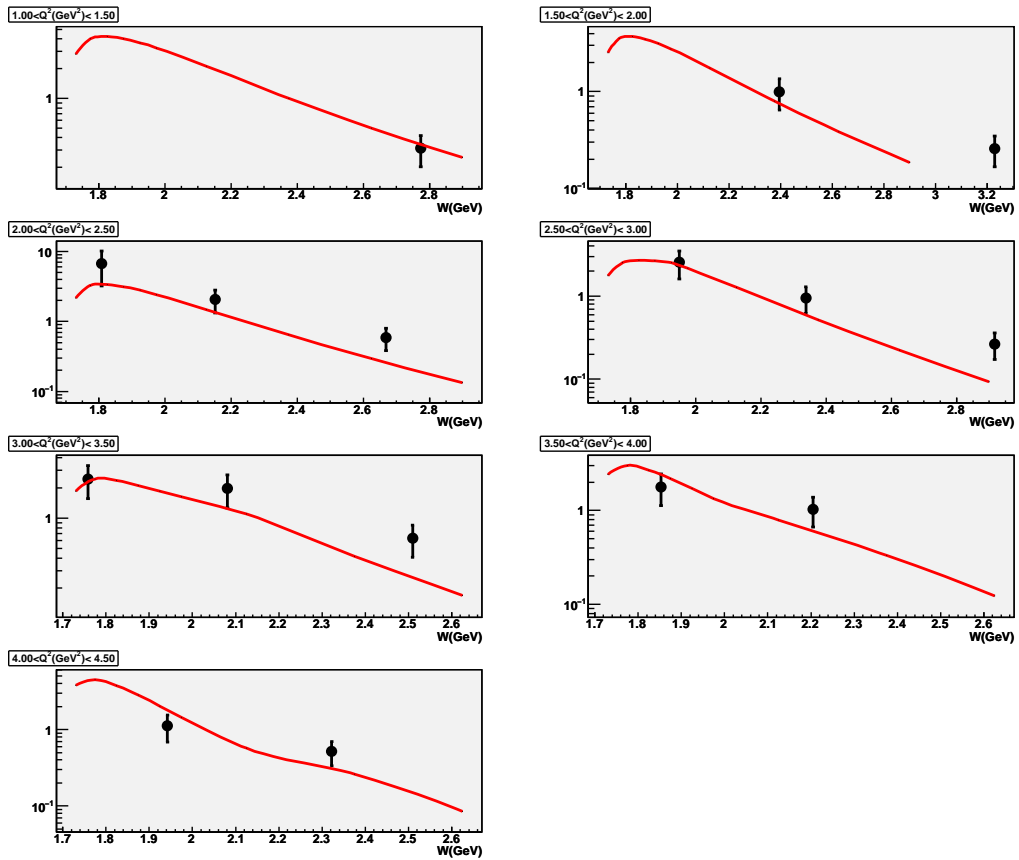}
\caption{\underline{PRELIMINARY} total cross section as a function of $W$ at fixed $Q^2$, for the reaction 
$\gamma^{*} p \to n \rho^+$. The red curves are the result of the JML model. Units are arbitrary on the y axis as the results are still
preliminary.}
\label{xsectionrhoPlaget}
\end{center}
\end{figure}

Figure \ref{xsectionrhoPlaget} shows the  total cross
section $\sigma(\gamma^* p \to n \rho^+)$ as a function of $W$ for fixed $Q^2$ compared with the JML calculation shown with the 
red curve. The JML model can successfully reproduce the cross sections for
almost all of  our $(Q^2,W)$ range. The decreasing  cross section with 
$W$ can be explained in the Regge theory by an intercept ($\alpha(0)=0.5$) of the $\rho^+$ trajectory which is dominant: $\sigma_{TOT} \propto
s^{\alpha(0)-1}$.

 \subsection{The GPD  ``partonic'' approach}

 The formalism of GPDs is valid in the so-called
Bjorken regime, i.e. $Q^2,\nu \to \infty$  with $x_B=\frac{Q^2}{2M\nu}$ finite. It was proven \cite{Collins97} that the dominant 
process for exclusive meson electroproduction, in the Bjorken limit, is given by
the so-called handbag diagram represented in Fig. \ref{reggeGPDs}. 
The handbag diagram is based on the notion of factorization in leading-order/leading-twist pQCD between a hard scattering process,
 exactly calculable in pQCD, and a nonperturbative nucleon structure part that is parametrized by the GPDs. For mesons, the factorization
  of the handbag diagram is only valid for the longitudinal part of the cross section.
In the following  we discuss the two particular GK \cite{GK} and VGG \cite{VGG} GPD-based calculations that provide
quantitative results for the longitudinal part of the exclusive meson cross
sections.

\begin{figure}[!h]
\begin{center}
\includegraphics[width=14cm]{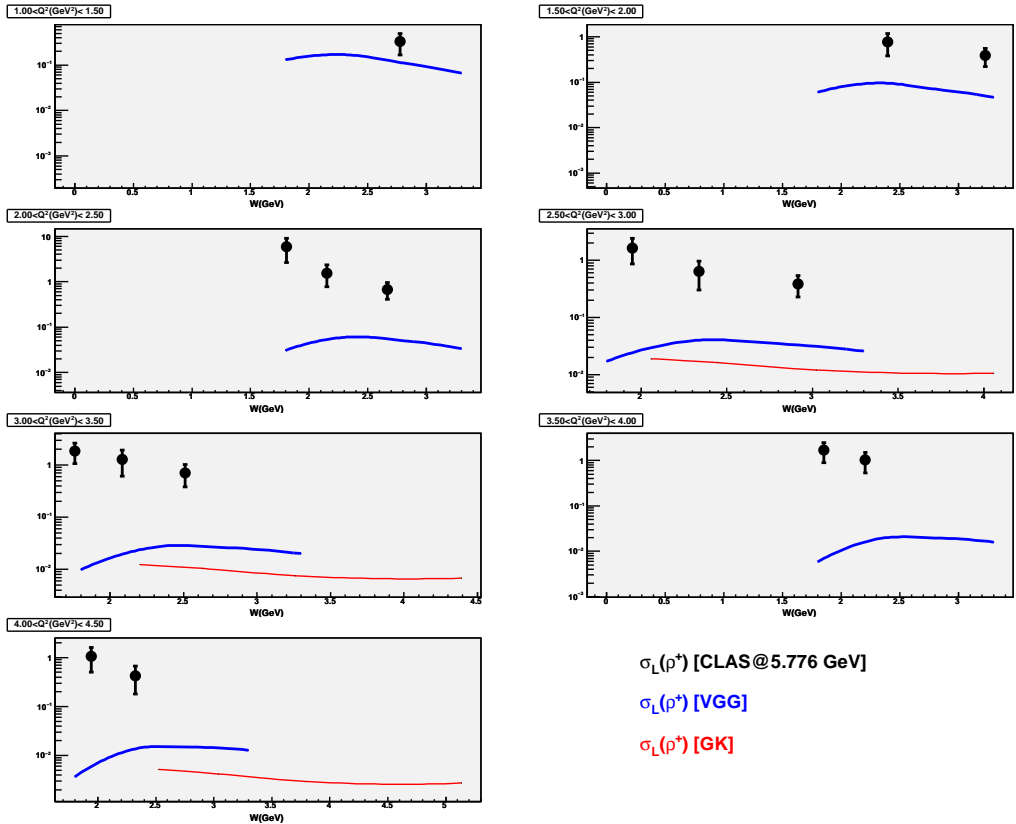}
\caption{\underline{PRELIMINARY} longitudinal cross section as a function of $W$ at fixed $Q^2$, for the reaction 
$\gamma^{*} p \to n \rho^+$. The red and the blue curves are the results of the GK and VGG models. Units are arbitrary on the y axis 
as the results are still preliminary. }
\label{theoryRhoplus}
\end{center}
\end{figure}

Figure \ref{theoryRhoplus} shows longitudinal cross section of $\gamma^* p \to n \rho^+$ as a function of $W$ and for fixed
$Q^2$.  The results of the calculations of the GK and VGG model are shown, respectively, with the red and the blue curves. We see
that both models fail to reproduce the data. This discrepancy can reach an order of magnitude at the lowest $W$
values. The trend of these particular GPD  calculations is to decrease as $W$ decreases, whereas the data increase. 
The same behavior was observed, in the low $W$ region, for the exclusive electroproduction of the $\rho^0$ \cite{rho}.

One can conclude that either the GPD formalism is not valid for our $Q^2$ coverage (due to important higher twist contributions for 
instance) or there
is a fundamental contribution, which is missing in the way the GPDs are parametrized in the GK and VGG models. 

\subsection{Comparison of the $t$ slope for the $\rho^+$, $\rho^0$, $\omega$ and $\phi$  channels}

The CLAS collaboration has measured  the largest ever set of data for the exclusive
 electroproduction on the proton of the vector mesons  $\rho^0$ \cite{rho}, $\omega$ \cite{omega} and $\phi$ \cite{phi}.
 
\begin{figure}[h]
\begin{center}
\includegraphics[width=16cm]{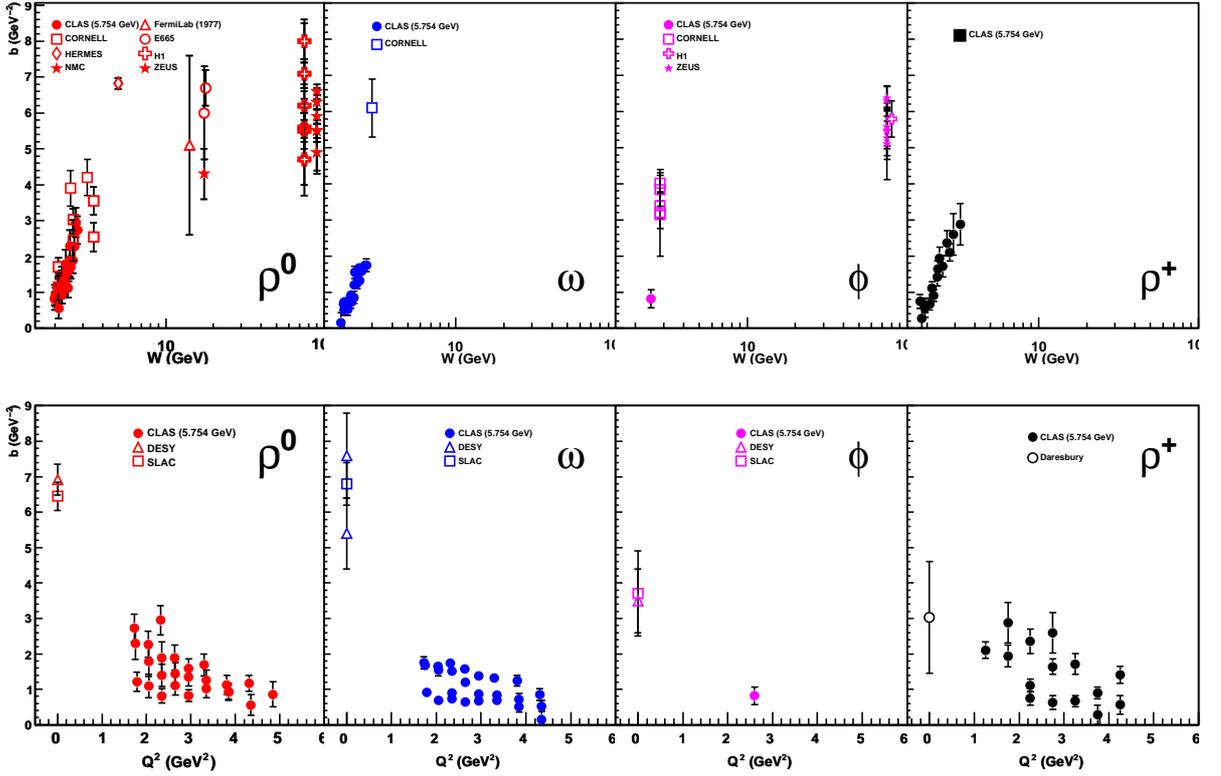}
\caption{The slope $b$ as a function of $W$ (on the top) and as a function of $Q^2$ (on the bottom) for the $\rho^0$, $\omega$, $\phi$
and $\rho^+$ channels.}
\label{tslopes}
\end{center}
\end{figure}

Figure \ref{tslopes} shows the slope of the differential cross section $d\sigma/dt$ for the  $\rho^+$, $\rho^0$, $\omega$ and $\phi$  channels as a 
 function of $W$ (in the top part)
 and as a function of $Q^2$ (in the bottom part). One can see the same trends of this slope, $b$, for all mesons channels, which can be interpreted
in simple and intuitive terms in the following way: 
 \begin{itemize}
 \item $b$ increases with $W$: the size of the nucleon increases as one probes the high $W$ values (i.e. the sea
 quarks), which could mean
 that the sea quarks tend to extend to the periphery of the nucleon.
 \item  $b$ decreases with $Q^2$: as we go to large $Q^2$, the resolution of the probe increases and we tend to see smaller and
 smaller objects.
\end{itemize}

\section{Summary}

We have presented (preliminary)  first ever measurement of 
 cross sections for the exclusive electroproduction of  $\rho^+$ on the proton using the CLAS detector at JLab.
These cross sections  can be interpreted with two approaches:
 \begin{itemize}
   \item hadronic approach: the JML model describes well most of the features of the $\rho^+$ cross
sections  up to $Q^2 \sim 4.5$ GeV$^2$.
 \item partonic approach: GPD models fail to describe longitudinal $\rho^+$ cross sections especially for low $W$.
 \end{itemize}
We also found the same trends of the variation of the $t$
slope as a function of $W$ and as a function of $Q^2$ for all  $\rho^+$, $\rho^0$, $\omega$ and 
$\phi$ channels.


\begin{thebibliography}{9}
\bibitem{clas} B.A. Mecking et al., 
Nucl. Instr. Meth. A {\bf 503}, (2003) 513 .
\bibitem{mathese} A. Fradi, thesis Univ. Paris-Sud at Orsay, 2009.
\bibitem{JML}
J.-M. Laget, Phys. Lett. B {\bf 489} (2000) 313.;\\
F. Cano and J.-M. Laget, Phys. Rev. D {\bf 65} (2002) 074022. ;\\
J. M. Laget, Phys.\ Rev.\ D {\bf 70} (2004) 054023.;\\
F. Cano and J. M. Laget, Phys.\ Lett.\ B {\bf 551} (2003) 317.
\bibitem{Collins97} J.C. Collins, L. Frankfurt and M. Strikman, Phys.Rev.D {\bf 56} (1997) 2982.
\bibitem{GK} S.V. Goloskokov and P. Kroll, Eur. Phys. J.C {\bf 42} (2005) 281; Eur.Phys. J.C {\bf 50} (2007) 829. 
\bibitem{VGG}
M. Vanderhaeghen, P.A.M. Guichon, M. Guidal, Phys. Rev. D {\bf 60}, 094017 (1999).
\bibitem{rho} S. Morrow et al., Eur.\ Phys.\ J.\ A {\bf 39}, (2009) 5.
\bibitem{omega} L. Morand et al., Eur.\ Phys.\ J.\ A {\bf 24} (2005) 445.
\bibitem{phi} J. P. Santoro  et al., Phys.\ Rev.\ C {\bf 78} (2008) 025210.
\end{thebibliography}
\end{document}